\documentclass[prl,twocolumn,superscriptaddress,a4paper]{revtex4}
\usepackage{graphicx}
\usepackage{epstopdf}
\usepackage{amssymb}
\usepackage{epsf}

\newcommand{\braket}[1]{\langle#1\rangle}

\usepackage{amssymb}
\usepackage{amsmath}

\begin{document}

\title{Phase sensitive measurements of order parameters for ultracold atoms \\
through two particles interferometry}

\author{Takuya Kitagawa}

\affiliation{Physics Department, Harvard University, Cambridge,
MA 02138, USA}

\author{Alain Aspect}

\affiliation{Laboratoire Charles Fabry de l'Institut d'Optique,
CNRS and Universit\'e Paris Sud 11, Campus Polytechnique, Avenue Augustin Fresnel, 91127 Palaiseau
cedex, France}

\author{Markus Greiner}

\affiliation{Physics Department, Harvard University, Cambridge,
MA 02138, USA}

\author{Eugene Demler}

\affiliation{Physics Department, Harvard University, Cambridge,
MA 02138, USA}

\begin{abstract}
Nontrivial  symmetry of order parameters is crucial in some of
the most interesting quantum many-body states of ultracold atoms and condensed matter systems.
Examples in cold atoms include $p$-wave
Feshbach molecules and $d$-wave paired states of fermions that could be
realized in optical lattices in the Hubbard regime.
Identifying these states in experiments requires measurements of
the relative phase of different components of the entangled pair wavefunction.
 We propose and discuss two schemes for such phase
sensitive measurements, based on two-particle interference revealed in atom-atom or atomic density correlations.
Our schemes can also be used for relative phase measurements for non-trivial particle-hole
order parameters, such as $d$-density wave order.
\end{abstract}
\date{\today}

\maketitle



The concept of order parameter, which characterizes states with
spontaneously broken symmetries, has been successfully applied to a
wide range of physical phenomena such as the Higgs mechanism in high
energy physics\cite{weinberg}, superfluidity in neutron
stars\cite{neutronstar}, superconductivity\cite{schrieffer},
gaseous Bose-Einstein condensates\cite{stringarirmp}  and
charge and spin ordering in electron systems\cite{kivelson}. Recent
works on condensed matter systems emphasized that order parameters
can often be characterized by non-trivial orbital symmetries. For example,
in contrast to conventional superconductors, which have isotropic $s$-wave electron
pairing, high Tc cuprates exhibit $d$-wave pairings\cite{dwave1},
while superfluidity of $^3$He or superconductivity in Sr$_2$RuO$_4$ exhibit triplet
$p$-wave pairings\cite{leggettmaeno}.
Other examples of order parameters with non-trivial orbital symmetries
discussed in the literature are high angular momentum Pomeranchuk
instabilities of electron systems\cite{pomeranchuk_charge} and
unconventional charge and spin density wave states\cite{ddw}.
Despite of the interests of such exotic states, 
the experimental verifications of these states are yet a
challenging problem. Standard thermodynamic and transport properties
can be used to observe the modulation in the magnitude of
quasiparticle gaps, but not the change of the sign of the order
parameters\cite{dwave1}. Only phase sensitive experiments,
such as the observations of Josephson effects in corner SQUID
junctions\cite{vanharlingen} and $\pi$-ring tricrystal
experiments\cite{tsuei}, have been considered as the definitive proof
of the unconventional pairing for both cuprates and
ruthenates\cite{nelson}. In the case of states with anisotropic
charge and spin orderings, the lack of experimental tests is one of the
main reasons that their existence remains an open question.

During the last few years, a considerable progress has been achieved in
creating analogues of strongly-correlated electron systems, using
ultracold atoms in optical lattices (see refs. \cite{bdz}
for reviews). One of the most challenging problems, which could be
addressed in the future experiments, is the search for $d$-wave pairing
in the repulsive Hubbard model\cite{hofstetter}.
Realizations of other exotic states in cold-atom systems, such as  $d$-density wave states\cite{honercamp},
have been theoretically proposed.
These states are characterized by order parameters with non-trivial
angular dependence of the relative phase between the components
of the entangled wavefunction.
Hence, it is important
to understand how tools of atomic physics can be used to perform
 tests of such quantum many-body states of ultracold atoms\cite{gritsev2008}.

In this paper, we discuss a scheme for performing such phase sensitive measurements.
It is based on the analysis of atom-atom correlations resulting from
two-particle interference\cite{2particlesinterferometry}.
Our proposal builds on the theoretical ideas\cite{altman} of using noise-correlations in atomic density to
characterize many-body states, and on the experimental demonstration
to measure atom-atom correlations, or atomic density noise spectroscopy with ultracold atoms\cite{aspect-helium1, bloch-bosons,noise_jila,heliumpairs}.
This method should provide an unambiguous evidence for
non-trivial pairings, including
$p$- and $d$-wave\cite{hofstetter,wang,gurarie}, as well as for
non-trivial
particle-hole correlations such as in a $d$-density wave state\cite{ddw,honercamp}.
It should also allow the direct observation of
two particle coherence and nontrivial angular momentum of ultracold
diatomic molecules. For example, for $p$-wave Feshbach molecules realized in
JILA\cite{p_wave_jila}, our approach should distinguish between $p_x+ip_y $
and $p_x-ip_y $ states\cite{gurarie}.
%



We start by considering a Feshbach molecule, which consists of a
pair of atoms, and has the center of mass momentum equal to zero
\begin{eqnarray}
| \Psi_{\rm mol} \rangle = \int  \frac{\textrm{d}^{3} \mathbf{k}}{ (2\pi)^{3/2}} \,  \psi(\mathbf{k})\, c_{\mathbf{k}\uparrow}^\dagger
c_{-\mathbf{k}\downarrow}^\dagger \, | 0  \rangle \label{wvf1}.
\end{eqnarray}
The two atoms making up the molecule can be either bosons or fermions.
For concreteness, in this paper we focus on the case of two fermions in different
hyperfine states labeled  by $\sigma =
\uparrow \downarrow$, in analogy with states of a spin 1/2 particle.
Here $\psi(\mathbf{k})$ is the wavefunction of a molecule,
$c_{\mathbf{k} \sigma}^\dagger$ is a creation operator of a fermion atom in the state
with momentum $\mathbf{k}$ and hyperfine state  $\sigma$. The symmetry of $\psi(\mathbf{k})$
determines the nature of the paired state.
We assume that the
potential binding the two atoms is removed instantaneously and the
released atoms subsequently evolve as free particles. Experimentally this can
be achieved either by changing the magnetic field abruptly near a
Feshbach resonance or by applying an RF pulse\cite{noise_jila,p_wave_jila}.
The released pair of atoms are in a superposition of momentum $(\mathbf{k,-k})$ pairs
with  amplitudes $\psi(\mathbf{k})$. Our goal is to find a method to
measure the relative phases between $\psi(\mathbf{k})$ for different $\mathbf{k}$.

\begin{figure}[t!]
\begin{center}
\includegraphics[width = 8.6cm]{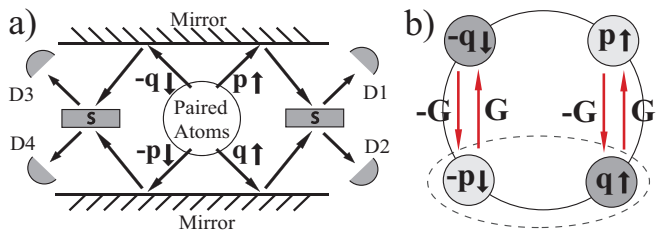}
\caption{Illustrations of using two-atom interference to measure the relative phase 
between different components of molecules after dissociation. 
 a) Scheme I:  Free propagating atoms are reflected in mirrors and mixed in beam splitters
denoted by S.
 Coincidences are counted between detectors on opposite sides, \emph{e.g.} $\textrm{D}_1$ and $\textrm{D}_3$.
 b) Bragg pulses with wave-vectors $\mathbf{G=p-q}$ and $\mathbf{-G}$
are used to exchange (atomic mirrors) or mix (atomic beam-splitters, S) components  $\mathbf{q}\uparrow$ and $\mathbf{p}\uparrow$,
 as well as $\mathbf{-q}\downarrow$ and $\mathbf{-p}\downarrow$($|\mathbf{q}| = |\mathbf{p}|$). In scheme II, the Bragg pulse
 applied at the beginning of expansion carries out reflections on mirrors and mixings in beam splitters in a single operation.
 }

\label{scheme}
\end{center}
\end{figure}
{\it Scheme I}. We first explain the main idea through the
scheme of Fig.\ref{scheme}a, analogous to the quantum optics scheme of \cite{2particlesinterferometrypapers}.
Atomic mirrors and beam splitters are used to reflect and mix states with momenta $ \textbf{p}$ and $ \textbf{q}$ on one side, 
$ \textbf{-p}$ and $ \textbf{-q}$ on the other side. Time and space resolved detectors in opposite sides (e.g. D1 and D3) allow 
measurements of correlations resulting from the interference between $\psi(\mathbf{p})$ and $\psi(\mathbf{q})$, 
and thus, can reveal the relative phase between these components.
The atomic mirrors and beam splitters are based on Bragg diffractions from laser beams which make standing waves with wave vectors $\pm \mathbf{G}$, and they couple states with the same spin and magnitude of momenta ($|\mathbf{q}| = |\mathbf{p}|$) but whose momenta differ by  $\mathbf{p-q}=\pm\mathbf{G}$. For simplicity, we consider
long Bragg pulses where a perfect Bragg diffraction can be achieved, i.e. no other diffraction order is involved. 
Later, we demonstrate that the phase sensitive measurements can also be carried out with short and
strong Bragg pulses which can introduce higher diffraction orders.
The atomic mirrors in
Fig.\ref{scheme} can be realized through Bragg pulses whose amplitudes and durations are chosen to produce $\pi$ pulse
so that it converts  $\pm\mathbf{p}$ into $\pm\mathbf{q}$ and vise versa. 
On the other hand, a $\pi/2$ pulse induces mixing between states with momenta $\pm\mathbf{p}$ and $\pm\mathbf{q}$, 
realizing a beam splitter (denoted by $S$ in Fig.\ref{scheme}). We can express the
original fermion operators in terms of the operators  after the mixing as follows:
\begin{eqnarray}
\mathrm{e}^{-\mathrm{i}\theta_{\mathbf{q} \uparrow}} \, c^\dagger_{\mathbf{q}\uparrow} &=& \cos \beta \, d_1^\dagger
-\mathrm{i} \sin \beta \, \mathrm{e}^{\mathrm{i}\chi_\uparrow} \, d_2^\dagger,
\nonumber\\
\mathrm{e}^{-\mathrm{i}\theta_{\mathbf{p} \uparrow}} \, c^\dagger_{\mathbf{p}\uparrow} &=&
- \mathrm{i} \sin \beta e^{-\mathrm{i}\chi_\uparrow}  \, d_1^\dagger + \cos \beta  \, d_2^\dagger,
\nonumber\\
\mathrm{e}^{-\mathrm{i}\theta_{\mathbf{-p}\downarrow}} \, c^\dagger_{\mathbf{-p}\downarrow} &=&  \cos \beta  \, d_3^\dagger
-\mathrm{i} \sin \beta e^{\mathrm{i}\chi_\downarrow}  \, d_4^\dagger,
\nonumber\\
\mathrm{e}^{-\mathrm{i}\theta_{\mathbf{-q}\downarrow}} \, c^\dagger_{\mathbf{-q}\downarrow} &=&
- \mathrm{i} \sin \beta e^{-\mathrm{i}\chi_\downarrow}  \, d_3^\dagger + \cos \beta  \, d_4^\dagger . \label{detectoroperators}
\end{eqnarray}
Here $d_i^\dagger$ are creation operators for particles observed in
detectors $\textrm{D}_i$ ($i=1,...,4$). The mixing angle $\beta$ (of the order of $\pi /2$) and spin-dependent phases
$\chi_\sigma$ can be controlled through the amplitudes, durations and relative phases of the Bragg laser pulses.
We denote by $\theta_{\mathbf{k}\sigma}$ the phase accumulated by an atomic component
with momentum $\mathbf{k}$ and spin $\sigma$ during the propagation between the source and the beam splitters.

If we assume that molecular wavefunctions for wave vectors $\mathbf{q}$ and
$\mathbf{p}$ differ only in  phase, \emph{i.e.} $\psi(\mathbf{k}) =
|\psi| \, \mathrm{e}^{\mathrm{i}{\phi}_\mathbf{k}}$, we find the following expressions
for the coincidence counts of $n_i=d_i^\dagger d_i$
\begin{eqnarray}
\langle \, n_1 \, n_3 \, \rangle_{\rm c} \,&=&\, |\psi|^2  \sin^2
(2\beta) \, \cos^2 \left( \frac{\phi_{\mathbf{q}}-\phi_{\mathbf{p}}+\Phi_{I}}{2} \right),
\label{conicidence_scheme1}
\nonumber\\
\langle \, n_1 \, n_4 \, \rangle_{\rm c} \,&=&\, |\psi|^2 \left[ 1-
\sin^2 (2\beta) \, \cos^2 \left( \frac{\phi_{\mathbf{q}}-\phi_{\mathbf{p}}+\Phi_{I}}{2}
\right) \right], \nonumber\\
\Phi_I &= &
\theta_{\mathbf{q}\uparrow} + \theta_{\mathbf{-q}\downarrow}  - \theta_{\mathbf{p}\uparrow} - \theta_{\mathbf{-p}\downarrow}  +\chi_\uparrow-\chi_\downarrow,
\end{eqnarray}
and similarly for $\langle \, n_2 \, n_3 \, \rangle_{\rm c}$
 and $\langle \, n_2 \, n_4 \, \rangle_{\rm c}$.
 The oscillatory behavior of the correlation as a function of $\Phi_I$
probes the coherence of pairing in the molecule.
To vary $\Phi_I$, one can, for instance,
change the phases  $\chi_{\sigma}$.
Moreover, if we know the precise value of $\Phi_{I}$, such coincidence signals yield
the relative phase $\phi_{\mathbf{q}}-\phi_{\mathbf{p}}$ between different molecular components.
In the absence of precise knowledge of $\Phi_I$,
the phase difference $\phi_{\mathbf{q}}-\phi_{\mathbf{p}}$
could be extracted through a scheme analogous to white light
fringes in classical optics, whose pattern and shape can reveal
the existence of fundamental phase factors\cite{whitelightfringes}.
Note, however, that $\mathbf{k}$ dependence of the phase factors acquired
during the propagation and the reflection may render these methods unreliable.
Thus, we consider a second scheme which avoids such a problem.

{\it Scheme II}.
In this alternative scheme, we apply a $\pi/2$ Bragg pulse
at the very beginning of the expansion to mix atomic components with 
momenta $\mathbf{q}\uparrow$ and $\mathbf{p}\uparrow$,  
as well as $-\mathbf{q}\downarrow$ and $-\mathbf{p}\downarrow$.
This realizes, in a single operation, reflections on the mirrors and mixing on
the beam splitters. 
In scheme II, there is a common mode propagation after the Bragg pulse, and 
phases acquired during the expansion do not affect interference.
Two atom interference is revealed by coincidence counts with point detectors just as in the previous scheme.
The scheme can be generalized to many-body case by replacing coincident counts
 between point detectors with density imaging and studying 
 noise correlations between patterns registered on opposite sides (see below).



To discuss scheme II, we start again  with the example of a dissociated  Feshbach molecule,  described by the
wavefunction in Eq.(\ref{wvf1}).
We consider the case in which the Bragg pulses  for spin up and down atoms
differ only in the phase, and such pulses are created by the potentials $ V_{\sigma}(\mathbf{r}) = 2V_0 \cos ( \mathbf{G.r} -
\chi_\sigma )$.
Here we assume again a perfect Bragg diffraction.
Detectors $\textrm{D}_i$ ($i=1,2,3,4$) detect atoms with momenta 
and spins $\mathbf{p}\uparrow$, $\mathbf{q}\uparrow$, $\mathbf{-q}\downarrow$,  $\mathbf{-p}\downarrow$, respectively.
The only difference between this scheme and scheme I is the absence of phase factors
$e^{\mathrm{i}\theta_{\mathbf{q}\sigma}}$.
As a result, coincidence counts have forms similar to 
Eq.(\ref{conicidence_scheme1}), with $\Phi_I$ replaced by $\Phi_{II}= \chi_\uparrow - \chi_\downarrow$.
Therefore, provided that we know the phase difference
$\chi_\uparrow - \chi_\downarrow$ associated with the two Bragg pulses,
atom-atom coincidence counts directly reveals
$\phi_\mathbf{q}-\phi_\mathbf{p}$, \emph{i.e.} the  pairing
symmetry in Eq.(\ref{wvf1}).

{\it Many-body state analysis}. We now apply scheme II
to a BCS state of fermions
$|\Psi \rangle = \prod_{\mathbf{k}} \, (u_{\mathbf{k}} +  v_{\mathbf{k}} {c}^\dagger_{\mathbf{k}\uparrow}
{c}^\dagger_{-\mathbf{k}\downarrow}) \,| 0 \rangle$.
This BCS wavefunction is general and can 
describe weakly-coupled BCS paired states as well as a condensate of
tightly-bound molecules.
Here, we consider the generic diffraction pulse that can mix states whose momenta are separated by any
integer multiple  of  $\mathbf{G}$. 
The effect of the mixing pulse is described by the transformation of
particle creation operators: ${c}^\dagger_{\mathbf{k}\uparrow}\,\rightarrow \,
\tilde{c}^\dagger_{\mathbf{k}\uparrow}=\sum_m \alpha^{\mathbf{k}\uparrow}_{0,m}
e^{-i m\chi_\uparrow} c^\dagger_{\mathbf{k}+m\mathbf{G}\uparrow}$, $
{c}^\dagger_{\mathbf{k-G} \uparrow}\,\rightarrow \,\tilde{c}^\dagger_{\mathbf{k-G} \uparrow}=\sum_m
\alpha^{\mathbf{k}\uparrow}_{-1,m}  e^{-i (m+1) \chi_\uparrow}
c^\dagger_{\mathbf{k}+m\mathbf{G}\uparrow}$, and analogously for
${c}^\dagger_{-\mathbf{k}\downarrow}$ and $ {c}^\dagger_{-\mathbf{k+G}\downarrow}$.
The scattering amplitudes $\alpha^{\mathbf{k}\sigma}_{j,m}$ are controlled by
the diffraction pulse amplitude $V_{0}$ and its duration $\tau$.
We assume that before the mixing pulse, only
states with momenta $\pm \mathbf{k}$, $\pm (\mathbf{k-G})$, which are close to the Fermi surface,
have finite probabilities to be occupied, while
states with momenta $\pm (\mathbf{k}-m\mathbf{G})$ for $m \neq 0, 1$ , which are far from the Fermi surface, are empty.
The mixing pulse then induces the interference between particles with momenta
$\pm \mathbf{k}$ and $\pm (\mathbf{k-G})$.


The signature of non-trivial pairing of the BCS wavefunction shows up in
the angular dependence of the phase $\phi_{\mathbf{k}}$ in $v_{\mathbf{k}}= |v_{\mathbf{k}}|e^{\mathrm{i}\phi_{\mathbf{k}}}$.
In order to probe the relative phase  $\Delta \phi = \phi_{\mathbf{k}} - \phi_{\mathbf{k-G}} $ between  pairs with momenta $\mathbf{k}$
and $\mathbf{k-G}$,
we consider the following density noise correlation after the interference:
\begin{eqnarray}
& \langle \delta n_{\mathbf{k}\uparrow} \, \delta n_{\mathbf{-k+G}
\downarrow} \rangle = \braket{n_{\mathbf{k}\uparrow} n_{\mathbf{-k+G}\downarrow}} -
\braket{n_{\mathbf{k}\uparrow} } \braket{ n_{\mathbf{-k+G}\downarrow}} \nonumber\\
&    = \left| v_{\mathbf{k}}u_{\mathbf{k-G}}
\alpha_{00}^{\mathbf{k}\uparrow}\alpha_{01}^{-\mathbf{k}\downarrow}
e^{-i\chi_{\downarrow}} + u_{\mathbf{k}}v_{\mathbf{k-G}}\alpha_{-10}^{\mathbf{k}\uparrow}
\alpha_{11}^{-\mathbf{k}\downarrow}e^{-i\chi_{\uparrow}}  \right|^2
\nonumber\\
&
  -\left( |v_{\mathbf{k}}|^2 - |v_{\mathbf{k-G}}|^2 \right) \times
\nonumber\\
& \left( |v_{\mathbf{k}}|^2 |\alpha_{00}^{\mathbf{k}\uparrow}|^2
|\alpha_{-10}^{\mathbf{k}\uparrow} \alpha_{11}^{-\mathbf{k}\downarrow}|^2  -
|v_{\mathbf{k-G}}|^2 |\alpha_{01}^{-\mathbf{k}\downarrow}|^2
|\alpha_{11}^{-\mathbf{k}\downarrow}|^2 \right).
\label{coincidence_eq}
\end{eqnarray}
In analogy with the case of a Feshbach molecule,
the first line in the RHS of Eq.(\ref{coincidence_eq}) contains an
interference term which depends on the relative phase $\Delta \phi$ as well as on
$\Phi_{II}= \chi_\uparrow - \chi_\downarrow$.

\begin{figure}[t!]
\begin{center}
\includegraphics[width = 8.6cm]{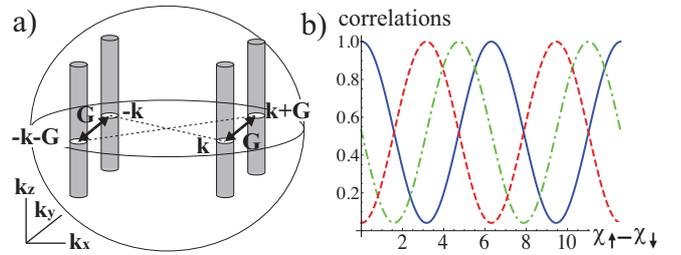}
\caption{a) In order to take into account the finite resolution of detectors and 
the integration in absorption imaging,
the density noise correlation is integrated over the cylinders shown
in the figure.
b) Integrated density noise correlations
$\langle \delta N_{V_\uparrow} \,\delta N_{V_\downarrow} \rangle $
as a function of phase difference $\chi_{\uparrow}- \chi_{\downarrow}$
for a strong diffraction pulse of amplitude $V_{0}/E_R=2$ and a duration $\tau$ which yields the maximum oscillation of the signal.
Blue, Green(dash-dotted line) and Red(dashed line) curves correspond to
$\Delta \phi =\phi_{\mathbf{k}} - \phi_{\mathbf{k-G}} = 0,\pi/2$ and $\pi$, respectively.}
\label{fig_cylinder}
\end{center}
\end{figure}

Space- and time-resolved single atom detection\cite{aspect-helium1,heliumpairs, single particle optical detection} 
permits direct measurements of atom-atom correlations for specific
momenta, corresponding to Eq.(\ref{coincidence_eq}).
Alternatively, one may look for noise correlation in absorption images after time of flight\cite{altman}.
In this case, absorption imaging, as well as finite resolution of detectors, result in
the integration of the atomic density. 
In order to take into account these effects, we have integrated Eq.(\ref{coincidence_eq})
over ranges of momenta as shown in Fig.\ref{fig_cylinder}a.
We present in Fig.\ref{fig_cylinder}b the numerical result of this integration, which displays noise correlation
 in integrated density \emph{vs.}  the phase difference
$\chi_{\uparrow}- \chi_{\downarrow}$ of the diffraction pulses.
Here we took the integration range to be $|\Delta k_{y}| = |{\bf G}|/10$, $|\Delta k_{x}| = |{\bf G}|/10$,
$|\Delta k_{z}| = 5|{\bf G}|$ and the pairing gap to be $\Delta \approx 0.1 E_{F}$.
The  diffraction pulse amplitude is set to $V_{0}/E_{R}=2$ where $E_R=|{\bf G}|^2/8m$ is the recoil energy, 
and its duration is chosen to have the maximum oscillation of the signal.
We assume that the integration range is sufficiently small that
the phases of the Cooper pairs $\phi_{\mathbf{k}}$ and $\phi_{\mathbf{k-G}}$ are constant
in the integration range.

\begin{figure}[t!]
\begin{center}
\includegraphics[width = 8.6cm]{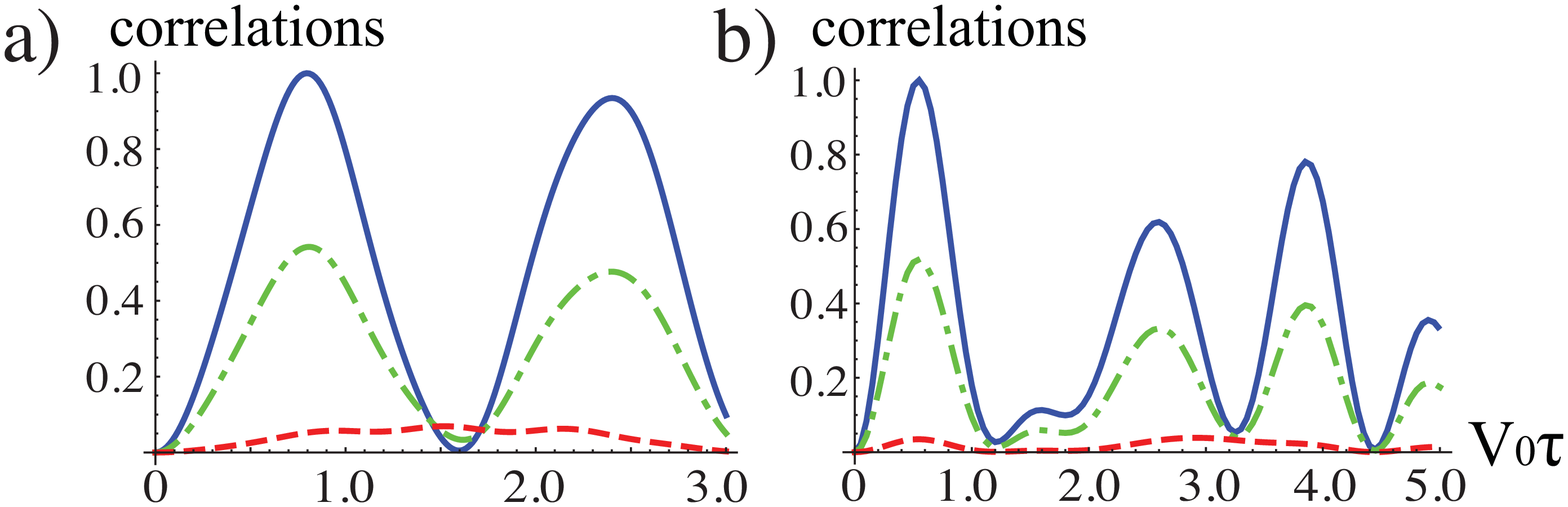}
\caption{Integrated noise correlations $\langle \delta N_{V_\uparrow} \, \delta
 N_{V_\downarrow} \rangle $ as a function of $V_{0}\tau$
 a) for a Bragg pulse amplitude $V_{0}/E_R=2$, and b) for
 a Bragg pulse amplitude $V_{0}/E_R=20$.
 Blue, Green(dash-dotted line) and Red(dashed line) curves correspond to
$\Delta \phi =\phi_{\mathbf{k}} - \phi_{\mathbf{k-G}} = 0,\pi/2$ and $\pi$, respectively.}
\label{weak_pulse_correlation}
\end{center}
\end{figure}

The oscillatory behavior of the integrated noise correlations $\langle \delta
N_{V_\uparrow} \, \delta N_{V_\downarrow} \rangle$ as a function of
$\chi_{\uparrow}- \chi_{\downarrow}$(See Fig \ref{fig_cylinder}b)
should provide an unambiguous
proof of the Cooper pair coherence. Moreover,  the value of the correlation at $\chi_{\uparrow}- \chi_{\downarrow}=0$ yields information
about  the phase difference $\Delta \phi= \phi_{\mathbf{k}} - \phi_{\mathbf{k-G}}$, which is the quantity we are interested in.
The value of the correlation at $\chi_{\uparrow}- \chi_{\downarrow}=0$ also depends on
the scattering amplitudes $\alpha^{k \sigma}_{j,m}$, and thus on $V_{0} \tau$.
In Fig.\ref{weak_pulse_correlation}, we present the integrated noise correlation signal $\langle \delta
N_{V_\uparrow} \, \delta N_{V_\downarrow} \rangle$
at $\chi_{\uparrow}- \chi_{\downarrow}=0$ as a function of $V_0 \tau$ for three different values of $\Delta \phi$, and find striking differences. We conclude that it should be possible to discriminate between  $\Delta \phi = 0$ and  $\Delta \phi = \pi$
even when full 3D resolution is not available.

In discussions so far, we assumed that the BCS pairs or
molecules are at rest before dissociation. When molecules are cold but not  condensed, there is a
spread in the center of mass momenta  determined by the
temperature. Even in this case, there is still a perfect coherence between
different parts of the wavefunction of each molecule, yielding a two-body interference.
However, the average of these interference terms over the center of
mass momenta of individual molecules could potentially result in the washing out of
noise correlations. We expect this suppression to be moderate as long as the
average thermal momenta of molecules is  smaller than the
characteristic momenta of expanding atoms.


{\it Systems with particle-hole correlations}. There are several
types of many-body states characterized by correlations in the
particle-hole channel, such as charge and spin
density wave states (CDW and SDW). The most exotic of them have
a finite angular momentum. This means that we have $ \langle
c_{\mathbf{k} \sigma}^\dagger c_{\mathbf{k+Q} \sigma} \rangle = \psi_{ph}(\mathbf{k}) $, where
$\psi_{ph}(\mathbf{k})$ has a non-trivial angular dependence.
\begin{figure}[t!]
\begin{center}
\includegraphics[width = 7.5 cm]{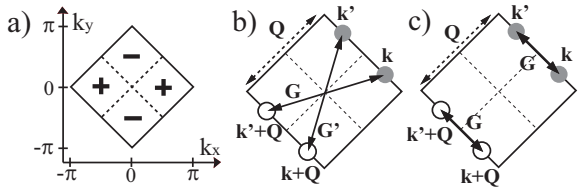}
\caption{a) Illustration of the sign of the CDW amplitude
$ \psi_{ph}(\mathbf{k}) = \langle c_{\mathbf{k} \sigma}^\dagger c_{\mathbf{k+Q} \sigma} \rangle  $
for {\it d}-wave CDW state.
 b), c) Phase sensitive detection of the symmetry of the {\it d}-wave CDW state in TOF experiments.
 For b), two pulses which transfer momenta $\mathbf{G}$ and $\mathbf{G'}$
 are applied at the beginning of expansion.
  In c), a single pulse with momentum transfer $\mathbf{G}$ is applied.
  All the couplings through the Bragg pulses are indicated by solid arrows.
 Here $\mathbf{Q}$ is the wave vector of CDW. }
\label{figure_dw}
\end{center}
\end{figure}
Our scheme above can be generalized to provide
an unambiguous phase sensitive detection of such states as well. To be concrete,
let us consider a 2D system near half-filling.
In this case, one can combine two different measurements of correlation functions to obtain the information on
the order parameter $\psi_{ph}(\mathbf{k})$, as shown in Fig.\ref{figure_dw}b and c. In Fig.\ref{figure_dw}b, two Bragg pulses  couple $\mathbf{k}$ and $\mathbf{k'+Q}$, as well as $\mathbf{k'}$ and $\mathbf{k+Q}$. Here, the correlation function $\langle \delta n_\mathbf{k} \delta
n_{\mathbf{k'}} \rangle $ contains an interference term proportional to
$\psi_{ph}(\mathbf{k}) \psi_{ph}(\mathbf{k'})$. In Fig.\ref{figure_dw}c, a Bragg pulse  couples $\mathbf{k}$ and $\mathbf{k'}$, and the correlation function $\langle \delta n_\mathbf{k} \delta n_{\mathbf{k'+Q}} \rangle $ contains the term $\psi_{ph}^{*}(\mathbf{k}) \psi_{ph}(\mathbf{k'})$.
When combined, these information
should not only provide  evidence of the angular dependence of CDW,  but also allow one to
distinguish site and band centered density wave states.


{\it Conclusion}.  In this paper, we have proposed a new method for performing phase sensitive measurements of non-trivial order parameters in systems of ultra-cold atoms, with a view toward  studying open problems in strongly correlated systems.
Note that in contrast to scheme I, which was introduced in analogy to a scheme first introduced in Photon Quantum Optics, scheme II is specific to Quantum Atom Optics, and takes advantage of the unique features of cold atom systems.

We acknowledge useful discussions
with E. Altman, I. Bloch, M. Lukin, and thank C.I. Westbrook for his suggestions on the manuscript. This work was supported by the
NSF grant DMR-0705472, Harvard MIT CUA, DARPA OLE program, AFOSR MURI, CNRS, ANR, Triangle de la Physique, IFRAF.

\end{document}